# Orientational frustration drives enhanced diffusion of anisotropic particles in a liquid labyrinth

Rohit Mangalwedhekar[1,2], Limeng Ruan[1,2], Somen Nandi[1,2], Quentin Gresil[1,2], Marc Tondusson[1,2], Stephane Bancelin[1,2], Lea-Laetitia Pontani[3,4], Laurent Cognet[1,2,*]

[1] *Laboratoire Photonique Numerique et Nanosciences, Université de Bordeaux, 33400 Talence, France,*
[2] *LP2N, Institut d'Optique Graduate School, CNRS UMR 5298, 33400 Talence, France*
[3] *Sorbonne Université, CNRS, Laboratoire Jean Perrin, LJP, F-75005 Paris, France*
[4] *Sorbonne Université, CNRS, Inserm, Institut de Biologie Paris-Seine, IBPS, F75005 Paris, France*

**ABSTRACT**

Transport of nanoscale objects in complex, structured environments plays a key role in a wide range of processes, from biomolecular dynamics in extracellular spaces to transport in porous materials such as filters and catalysts. While anomalous diffusion is well established, how particle anisotropy governs transport under geometric constraints remains unclear. Here we use 3D single-particle tracking to investigate the diffusion of stiff one-dimensional carbon nanotubes in a continuous soft matter network of interconnected chambers and constrictions. Transport is anomalous and antipersistent, with strong length dependent confinement and trapping, consistent with obstructed diffusion. Unexpectedly, however, escape from confinement is poorly sensitive to nanotube length as opposed to what would be expected of pore mediated transport. Despite a tenfold length increase and significantly enhanced trapping, escape time increased by only ~1.4×. Single-particle orientational tracking reveals the origin of this weak scaling. Indeed, long nanotube, i.e. those with length comparable to the chamber dimensions, dynamically align with constrictions enabling efficient, geometry-assisted escape that offsets increased confinement while shorter nanotubes need to screen the volume to find their escape path.

These results uncover an alignment-mediated transport mechanism that decouples confinement strength from escape kinetics, distinct from pore-mediated transport mechanisms, establishing a quantitative framework for anisotropic diffusion in complex environments.

## Introduction

Diffusive processes in natural systems are generally described within frameworks that explain how stochastic motion evolves over time. While Brownian motion provides the simplest model for such processes[1,2], real-world environments often exhibit significant deviations from simple Brownian statistics[3]. The transport of nanoscale objects in many natural systems that have complex structures and confinements display such deviations from Brownian mechanics[4]. These processes range from biomolecular activity within cells or intercellular communications in the extracellular spaces (ECS) of tissues to the diffusion of fluids through filters, catalysts, and metalorganic frameworks and pose critical challenges and opportunities both for fundamental research and technological innovation[5–13].

In recent years, diffusion in confined and complex environments has therefore received widespread attention. This growing interest has been significantly advanced by breakthroughs in single-particle tracking (SPT) [14,15] and the development of environmental models as well as diverse fluorescent probes, which can passively or actively navigate confined spaces[7,16–22]. When applied to biological systems, single-molecule diffusion measurements have revealed anomalous diffusion patterns driven by local variations in viscosity, biomolecular crowding, and intrinsic fluxes within cellular and tissue architectures [6,23–27]. Understanding diffusion in such intricate environments, and unravelling the origins of anomalous behavior, is not only a fundamental scientific challenge but also essential toward advancing our knowledge of health, disease and treatment strategies.

Much of our current understanding of diffusion in complex systems has been established using spherical probes, such as small fluorescent molecules, quantum dots, or fluorescent beads[22,26,28]. In synthetic environments such as porous gels, packed colloids, and inverse opals, these isotropic probes have been instrumental in demonstrating and characterizing how environmental heterogeneities give rise to anomalous behavior and non-Gaussian statistics[9,10,29–32]. In biological samples, diffusing entities often exhibit limited isotropy, leaving the influence of particle shape on transport, confinement, and escape mechanisms largely unexplored. Additionally, anisotropic particles may offer further insights by highlighting specific environmental features that govern diffusion, features that spherical particles might overlook[33]. Experiments on the free diffusion of anisotropic particles have demonstrated the coupling between orientational and translational diffusion[34].

In recent years, single-walled carbon nanotubes (CNTs) have successfully been used as single particle fluorescent probes to study diffusion in biological tissues. The combination of their bright and photostable luminescence in the short-wave infrared with their high 1D aspect ratio and tiny diameter indeed enables probing bioenvironments at depth inside preserved tissues. Notably, investigations of the brain ECS using CNTs of varying lengths ranging from 50 to several hundred nanometers have revealed anomalous diffusion while simultaneously enabling the mapping of intricate structures previously inaccessible to other techniques [16,35–37]. Recent work on the SPT of CNTs in the ECS of the brain has shown that hindered diffusion emerges as a result of geometry[38]. Similarly, CNTs exploring interstitial spaces in highly autofluorescent mouse liver tissues display pronounced variations in anomalous diffusion between healthy and fibrosed states[39].

Beyond biological tissues, the diffusion of rigid nanorods within crowded and cross-linked synthetic networks has been extensively studied in polymer physics. Notably, nanorods longer than the mesh size of entangled or cross-linked polymer networks have been shown to diffuse anomalously fast compared to the predictions of reptation or simple obstruction models, owing to a decoupling from polymer relaxation dynamics[40]. More recent work has directly linked this enhanced mobility to a coupling between rotational and translational dynamics, whereby nanorod reorientation facilitates passage through adjacent mesh openings[41], with the magnitude of the effect depending sensitively on nanorod length, rigidity, and surface roughness. These studies, conducted in mesh-like, effectively concave confining geometries, provide an important point of comparison for the present work, which instead probes a continuous, convex landscape of interconnected chambers and tunnels.

Long CNTs have been used to study the reptation and elastic behavior of semiflexible filaments across diverse diffusing environments[42,43]. However, reports on the dynamics of rigid anisotropic diffusers in confined spaces remain limited. While theoretical studies have explored the diffusive escape of rigid rods within idealized confined geometries[44], experimental investigations into the diffusion of anisotropic rigid particles in complex environments is still poorly understood. Within this context, how rigid anisotropic particle geometry interacts with local architecture to shape diffusion, orientation dynamics, confinement and escape statistics in soft-matter environments appears as a timely and key question. Beyond biological environments, this question is also relevant to molecular movements in zeolites and porous media [8–10,29–31]. Addressing this question calls for well-defined model systems that decouple geometrical effects from, e.g. chemical content or biological complexities, enabling controlled investigation of transport mechanisms.

In this work, we demonstrate how the interstitial spaces within densely packed emulsion droplets can act as a biophysical model for studying 3D single-particle diffusive processes in complex soft matter environments as an idealized proxy of the ECS freed from complex molecular crowding. These densely packed droplets have already been instrumental to serve as model systems to explore the macroscopic response of tissues, particularly the roles of mechanical forces, rheology, and cell–cell contact interactions[45–48]. The intricate network of interstitial spaces between these droplets provides a unique landscape for investigating the diffusion of single molecules at the nanoscale. By introducing diffusing anisotropic CNTs, we examine how geometric frustration of their rotational motion within continuous convex confinements enhance the translational diffusion and escape dynamics of rigid anisotropic particles.

Within these environments we demonstrate the anomalous 3D diffusion of anisotropic particles due to purely geometrical effects. We further show how local geometries influence diffusion of these anisotropic particles depending on their length. Importantly, we reveal that in such a convex landscape, the escape time of particles from local areas scales more weakly with particle length than expected in concave spaces. This effect is attributed to local alignment induced by frequent particle–wall interactions that pre-align particles and favor their escape. To validate this interpretation, we employ single-particle orientation tracking (SPoT), demonstrating how geometric frustrations differentially influence particle orientation depending on particle length. These interactions ultimately shape escape mechanisms and affect the scaling of escape times for anisotropic particles.

## Results

We selected a soft matter system composed of densely packed emulsion droplets (~5 μm in diameter) whose interstitial spaces form a network of continuous convex architecture representative of a complex soft matter diffusion landscape such as the brain ECS. This intricate network of interstitial space serves as our biophysical model system used to challenge current diffusion models in complex soft matter.

We used (6,5) CNTs functionalized with quantum color centers (called as CCNTs) emitting at ~1140 nm based on an aryl substituent as the functional group[16]. This functionalization yields bright nanotubes at typical lengths of ~500 nm (mean), and further enables the production of ultrashort luminescent nanotubes (uCCNTs), with a mean length of about 50 nm[16] as measured using Atomic Force Microscopy (AFM) (Supplementary Figure 1). Consequently, both CCNTs and uCCNTs can serve as highly effective probes for deep-tissue imaging at depths exceeding 100 μm, highlighting the transformative potential of anisotropic probes in revealing the complexities of biological structures. Imaging was performed using a custom-built single molecule setup equipped with a double helix phase mask to engineers the point spread function (PSF), enabling 3D single-particle tracking (SPT) of (u)CCNTs in the near-infrared[49,16]. This approach allowed us to record ultra-long trajectories of CCNTs and uCCNTs within the emulsion droplets (Figure 1a).

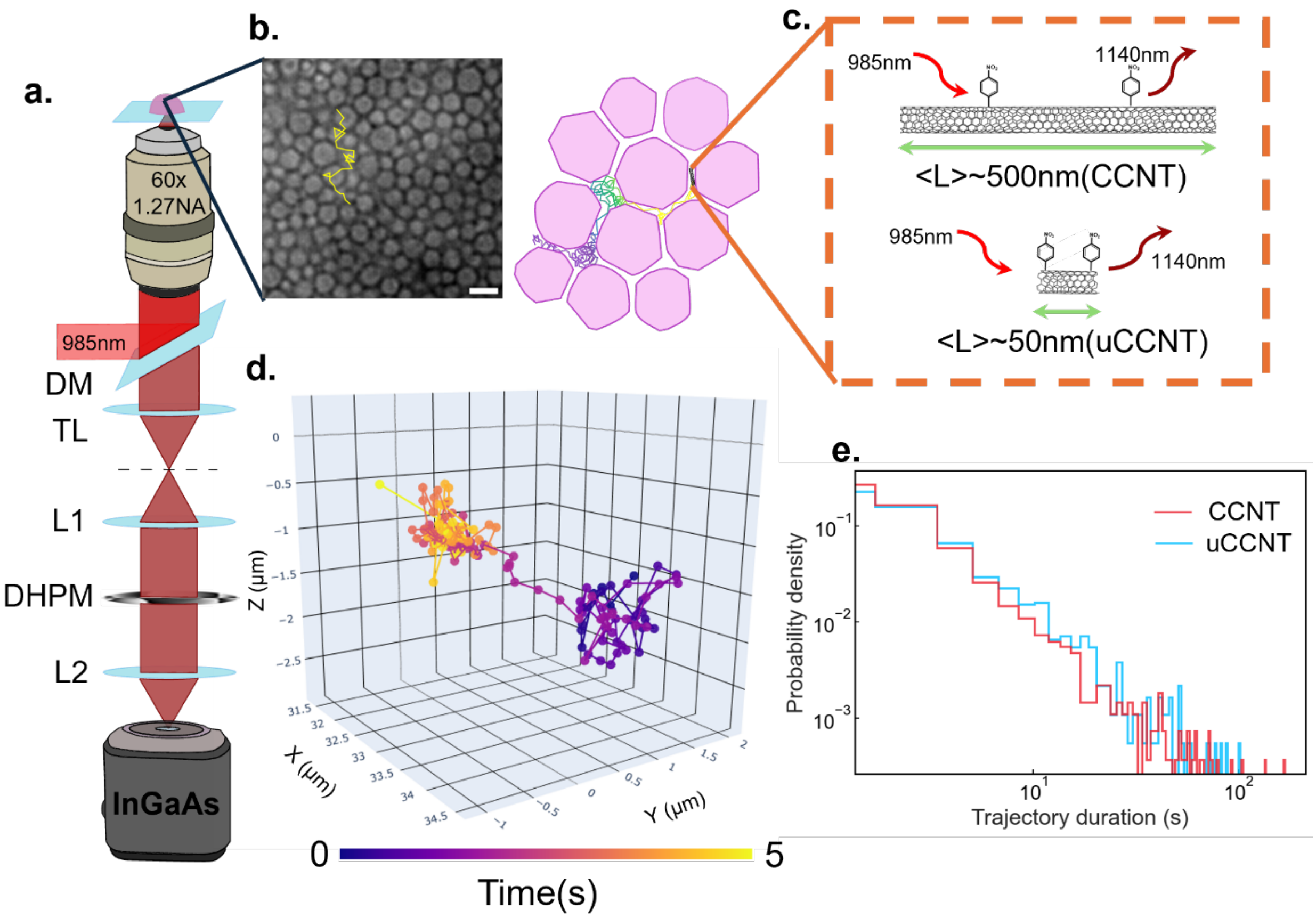

**Figure 1 3D Single Particle Tracking of individual Carbon nanotubes in a biophysical model** a. Schematic widefield microscopy setup equipped with a Double Helix Phase mask (DHPM) to achieve 3D single particle tracking of diffusing individual CNTs. b. An optically sectioned fluorescence image representative of the packed emulsion droplet structure as a biophysical model and a schematic representation of a CNT diffusing in interstitial regions. Scale bar: 5 μm. c. Schematic representation of the long CNTs (CCNTs) and ultra-short CNTs (uCCNTs) used to study the length dependence of diffusion in complex structures. Both types of CNTs were functionalized to fluoresce brightly with high photostability at ~1140 nm when excited at 985 nm. d. Typical 3D trajectory of a single CNT tracked while exploring the biophysical model. e. A histogram of acquired trajectory lengths. The median trajectory length is around 2s and the average trajectory length is about 5s for both CCNTs and uCCNTs.

The biophysical model exhibited a highly dense packing under white-light imaging (Figure 1b). To quantify its structure more precisely, we fluorescently labeled the droplet interfaces with Nile Red and implemented the HiLo modality (SPARQ, Bliq Photonics), which provides optical sectioning for imaging fluorescently tagged droplet surfaces[37,50] (see Droplets imaging under Materials and methods). Fluorescence images revealed a tight packing evidenced by the deformations of the individual droplets from nearly circular to polygon-like shapes[50]. The distribution of the packed droplets sizes (median diameter ~4.3 μm) was estimated using the Hough Circle Transform in Fiji (Supplementary Figure 2a) from the optically sectioned 2D fluorescence images. The fraction of the space occupied by the interstitial spaces between droplets was quantified by thresholding optically sectioned images in Fiji. This analysis revealed that interstitial regions account for approximately 12 ± 1% of the total sectioned area. Next, the biophysical model was prepared with the inclusion of (u)CCNTs (Figure 1c). A total of 2481 and 1564 trajectories of CCNTs and uCCNTs were acquired respectively, providing statistically robust interpretations. In both cases, the median trajectory length was ~2 s corresponding to ~60 data points (Figure 1d, e), ensuring comparable temporal sampling across particle lengths. Some individual CCNTs were also imaged for longer timescales (up to 160 s, i.e. 4874 data points) allowing to retrieve the network structure of the interstitial regions between the emulsion droplets.

Diffusion of anisotropic particles was first analyzed by calculating the ensemble Mean Squared Displacement (MSD - $< r^2(\Delta t) >$) to understand the scaling of exploration in the structure[51].

$$< r^2(\Delta t) > = K_\alpha (\Delta t)^\alpha \tag{1}$$

where $\alpha$ is the anomalous diffusion exponent, $\Delta t$ the lag time between two positions, and $K_\alpha$ the generalized diffusion coefficient.

The ensemble MSD exhibits clear sublinear scaling at short time lags, with diffusion exponents $\alpha = 0.75 \pm 0.02$ (uCCNTs) and $\alpha = 0.64 \pm 0.01$ (CCNTs), establishing that diffusion is anomalous for both particle lengths despite a tenfold size difference (Figure 2a). The generalized diffusion coefficients were determined as $K_\alpha = 1.43 \pm 0.01$ μm$^2$/s$^\alpha$ for uCCNTs and $K_\alpha = 0.70 \pm 0.01$ μm$^2$/s$^\alpha$. Next, we use the Velocity Autocorrelation Function (VACF) as a parameter that quantifies temporal correlations in particle displacements and reveals memory effects through interactions with the environment[51] (Figure 2b).

$$\tilde{C}_v(\Delta t) = \frac{\langle \boldsymbol{v}(t + \Delta t) \cdot \boldsymbol{v}(t) \rangle}{\langle \boldsymbol{v}(t)^2 \rangle} \tag{2}$$

where $\boldsymbol{v}(t)$ is the velocity component of the particle at time $t$ and $\Delta t$ is the time lag. The VACF calculated at the fourth time lag independently corroborates anomalous transport with antipersistent steps, displaying pronounced negative correlations from $\Delta t \approx 0.4 - 1.0$ s before returning to zero, with stronger antipersistence for CCNTs than for uCCNTs.

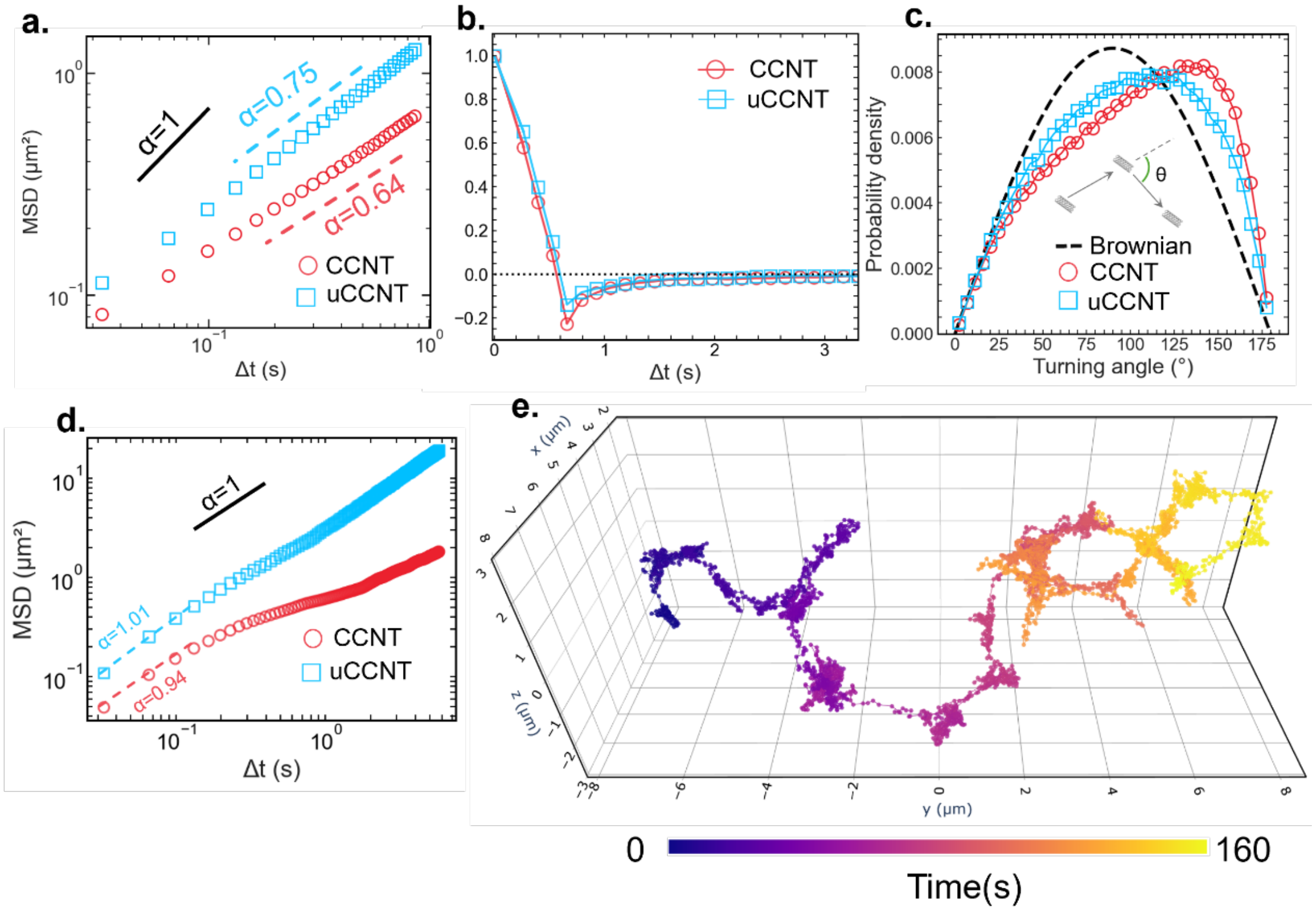

**Figure 2 Diffusion analysis of 3D SPT of nanotubes reveals anomalous behaviour** a. MSD versus lag time for CCNTs and uCCNTs in the biophysical model showing subdiffusive anomalous behaviour. b. The velocity autocorrelation function (VACF) exhibits a negative correlation, highlighting the restriction CNTs face when bouncing against the walls of the droplets. c. The right shifted nature of distribution of turning angle (measured as the exterior angles as shown in inset) confirms the increased interactions with the walls as compared to Brownian motion (black dotted line). d. MSD of single long trajectories of uCCNTs and CCNTs illustrating that within a single trajectory diffusion is complex and contains multiple regimes. e. Representations of the cartesian coordinates of a long single trajectory indicates the presence of local geometries of "chambers" and "tunnels".

Lastly, we choose to assess the distribution of turning angles as a direct probe of directional constraints in molecular motion[51] (Figure 2c). Here, we define the turning angle as the exterior angle between successive displacement vectors formed by three consecutive particles' localizations, quantifying the change in direction of motion independent of particle orientation (inset Figure 2c). The distribution from the trajectories of both the uCCNTs and CCNTs exploring the biophysical model revealed a right shift towards large angles with a maximum at 110° and 142° respectively, as compared to distribution of standard Brownian motion (dashed black line, Figure 2c). The systematic shift toward large turning angles reflects repeated wall-mediated deflections. The stronger shift observed for CCNTs demonstrates that increasing particle length amplifies geometric frustration and the frequency of boundary interactions (Figure 2c). Together, these observations strongly suggest that subdiffusive anomalous diffusion in the biophysical model arises from geometric frustration and are consistent with the obstructed diffusion model[38]. As a control, the MSD, VACF, and distribution of turning angles were also calculated for Brownian diffusion in a homogeneous imaging buffer solution (Supplementary Figure 3).

Some individual CCNTs were also imaged for longer timescales to retrieve the network structure of the interstitial regions of the biophysical model. While ensemble averaging smooths out these features, individual long-trajectory MSDs clearly resolve three distinct dynamical regimes characteristic of transient trapping (Figure 2d), exhibiting a near linear growth in the first regime ($\alpha$ = 0.94 and 1.01 for CCNTs and uCCNTs respectively), followed by sublinear or near plateau-like behavior in the second regime, and finally, near linear growth again indicating free diffusion at longer time scale (~1 s). Interestingly, direct examination of the particle trajectories in the interstitial spaces clearly demarcates different geometries explored by the particles such as voluminous "chambers", and channel like "tunnels" as will be discussed below (Figure 1d, 2e). Furthermore, these geometries were also visualized when the droplet environment was imaged with optical sectioning. This motivates a geometry-resolved analysis of particle motion as is discussed below.

To study the effect of local geometry on diffusion, the long trajectories were manually segmented into two main distinct topologies using the software CloudCompare: chambers and tunnels, connecting various chambers (Figure 3a). The MSD and turning angles of the particles in these two topologies were next studied separately. The MSD of uCCNTs in the chambers grow almost linearly before turning sublinear, indicating the time required for the nanotubes to explore their local environment. Within chambers, uCCNTs initially exhibit near-linear MSD growth before transitioning to subdiffusion at ~330 ms, consistent with rapid free volume exploration followed by confinement-limited motion. In contrast, long CCNTs display sublinear MSD growth at lower time scales, evolving into a plateau indicative of transient trapping at ~200 ms (Figure 3b).

In the chambers, the distribution of turning angles for both ultrashort and long CCNTs is close to Brownian at the earliest time lag (Δt=0.033 s). However, at longer time lags (Δt=0.33 s, Δt=0.825 s) the strong shifts progressing towards large turning angles indicate prominent interaction with walls (Figure 3c, d).

Notably, within tunnels, while MSDs alone suggest nearly Brownian transport for both particle lengths (Figure 3e), turning-angle statistics reveal fundamentally different underlying dynamics. For long CCNTs, the turning-angle distribution evolves from homogenous (Δt=0.033 s) toward bimodal form (Δt=0.825 s), with peaks at small angles (persistent motion) and large angles (antipersistence), reflecting aligned (forward) and anti-aligned (backward) steps along the tunnel axis (Figure 3f). Such a bimodality is absent for uCCNTs, which retain progressively antipersistent turning statistics at longer time lags (Δt=0.33 s, Δt=0.825 s) due to repeated collisions with the boundary walls while showing nearly Brownian distributions at very early time lags (Δt=0.033 s) (Figure 3g). These length-dependent turning statistics indicate that particle length determines how geometric frustration is resolved: uCCNTs undergo near isotropic exploration with frequent wall interactions, whereas long CCNTs adopt alignment-assisted transport along tunnels.

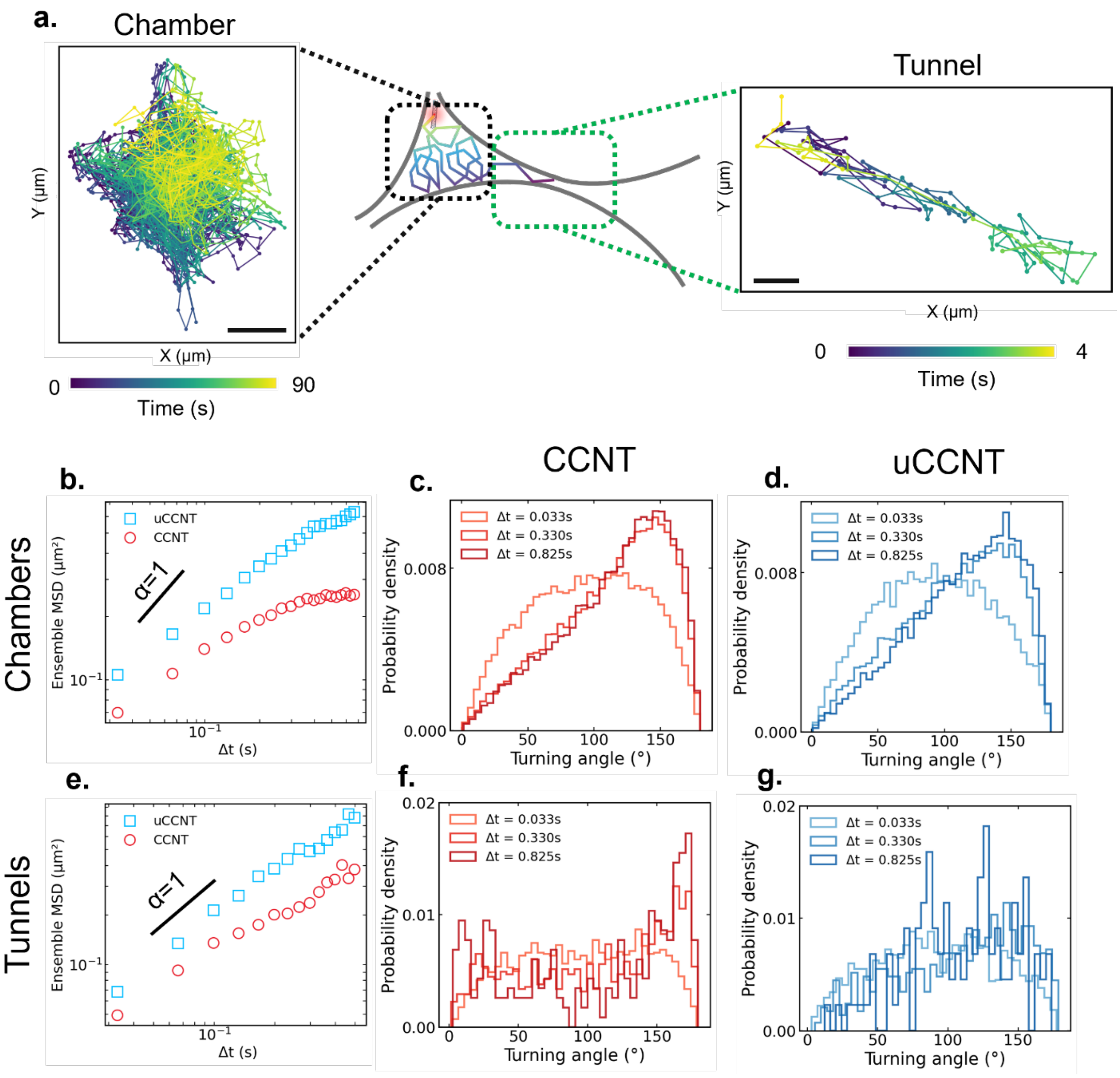

**Figure 3 Diffusion of nanotubes is influenced by local geometry** a. Schematic of a nanotube trajectory in the model system, displaying the two distinct environmental topologies, together with representative trajectories in chambers (Scale bar: 500 nm) and tunnels (Scale bar: 200 nm), manually segmented on CloudCompare. b. MSD versus lag time for CCNTs shows transient trapping in the chambers while the uCCNTs show strongly subdiffusive behaviour at longer lag times. c, d. Distribution of turning angles in the chambers shows near Brownian exploration in the first time lag but right shifted distributions at longer time lags, confirming strong subdiffusive behavior for both CCNTs and uCCNTs. e. MSD versus lag time for both CCNTs and uCCNTs reveals nearly Brownian diffusivity in the tunnel regions. f, g. Distribution of turning angles in tunnels is strongly parallel or anti-parallel with significant under-representation of intermediate angles owing to the geometric and steric constraints at higher time lags for long nanotubes. Conversely the ultra-short nanotubes continue to show right shifted angles at different time lags indicating obstructed yet Brownian diffusion.

Next, we studied the distribution of escape times, defined as the time taken by the nanotubes to exit a chamber. We determined the escape times, $\tau_{long}$ (for CCNTs) and $\tau_{short}$ (for uCCNTs), from 150 and 148 chamber-residency events respectively (Figure 4a, see Methods for extraction criteria and Supplementary Figure 2c for the resulting, overlapping chamber-size distributions of the two populations). Strikingly, despite a tenfold difference in nanotube length, we found the ratio $\tau_{long}/\tau_{short} \approx 1.4$ only. To estimate statistical uncertainty, we use a bootstrap resampling (N=100 000) that results in a 95% confidence interval range of [1.04, 1.85] with a median of 1.38 (Figure 4b).

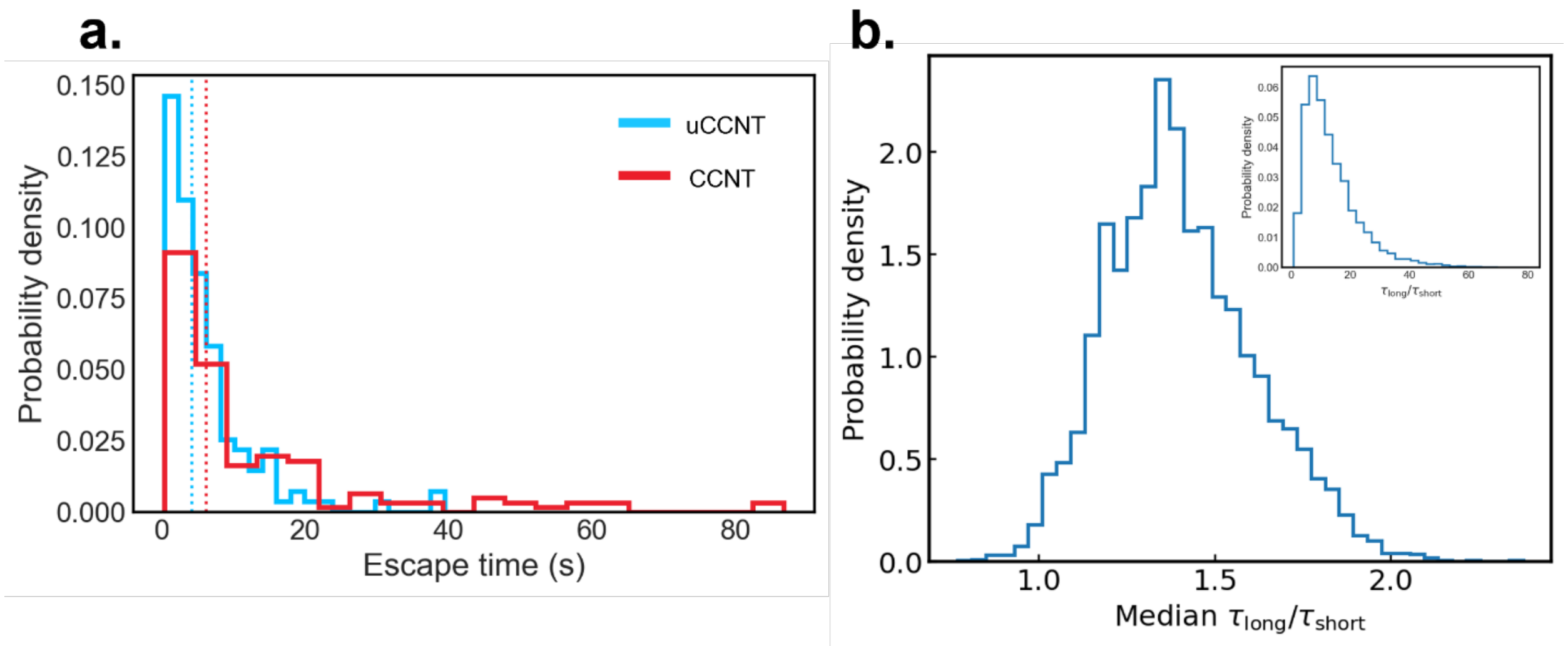

**Figure 4 Length dependence on escape times from chambers** a. The distribution of residency times of uCCNTs and CCNTs reveals a median escape time of around 3.3 s and 5 s, respectively. Thus, the escape time ratio for long and short particles is 1.38. b. A bootstrap resampling (N=100 000) was used to quantify the 95% confidence interval of the median escape time ratios to be [1.04, 1.85]. A Monte Carlo parameter sweep simulation was used to theoretically predict ratio of escape times by Brownian diffusion of anisotropic particles from confinement and the median value was found to be 11.28 ± 9.66 (inset).

To test whether this weak length-dependence is consistent with purely diffusive escape, we compared it to a quantitative model of Brownian escape for anisotropic particles from confinement (Doi and Xu model[44]), evaluated via a Monte Carlo parameter-sweep simulation built on our measured geometry (see Methods). This purely Brownian model predicts a linear scaling of escape time with particle length. For a tenfold increase in nanotube length, it predicts a median escape-time ratio of 11.28 ± 9.66 (Figure 4b, inset). This predicted ratio is much larger than the ~1.4-fold increase we measured experimentally. We examine the physical origin of this deviation in the Discussion.

The striking reduction in length sensitivity reveals that escape is not limited by translational or rotational diffusion alone. Instead, based on the length of the particle, we hypothesize the confinement induces a geometric frustration of rotational dynamics that adjusts escape mechanisms for the uCCNTs and CCNTs and compensate for the increased length of the particle. This reduced scaling of escape times along with the distribution of angles in the chambers and tunnels points to geometry-induced alignment of long anisotropic particles to the tunnels that mitigate rotational constraints and facilitate escape.

We next place this interpretation on a more quantitative aspect by characterizing the geometry of the biophysical model itself. For a statistically homogeneous and locally isotropic medium such as our densely packed droplet network, the area fraction measured on a random 2D section is, by the classical Delesse stereological principle, an unbiased estimator of the corresponding 3D volume fraction[52,53]. The interstitial area fraction reported above (12 ± 1%) therefore provides a direct estimate of the 3D porosity of the biophysical model, $\phi \approx 12 \pm 1\%$ (Table 1a in Supplementary Information).

Independently of this porosity estimate, the chamber and tunnel dimensions define the dimensionless confinement ratios (Table 1b in Supplementary Information): for CCNTs, $L/D_c \approx 0.6\text{-}0.8$ and $L/a \approx 4.5\text{-}7.1$, placing the long nanotube length within a factor of ~1.5 of the chamber size and several tunnel diameters. For uCCNTs, $L/D_c \approx 0.07$ and $L/a \approx 0.45\text{-}0.71$, an order of magnitude smaller on both confinement scales. These same quantities (chamber volume V, nanotube length L, tunnel radius a) are the direct inputs to the Doi-Xu escape-time model (Eqs. 3-6) and to the Monte Carlo parameter-sweep simulation below (Figure 4b inset): the theoretical comparison presented in this work is therefore calibrated directly by our own measured geometry.

The tunnel and radius distributions obtained independently from CCNT and uCCNT trajectories (Supplementary Figure 2d) further indicate that the escape mechanism is constrained on purely steric grounds: all the tunnels have a radius smaller than half the CCNT length (a < 250 nm), meaning CCNT passage requires substantial reorientation in virtually every escape event; by contrast, all of the same tunnels are wider than half the uCCNT length (a > 25 nm), consistent with the near-isotropic, alignment-free passage inferred for uCCNTs from the turning-angle statistics (Figure 3g). This purely geometric analysis provides independent quantitative support for the length-dependent escape mechanism proposed above.

To probe and confirm directly this orientational mechanism underlying length-dependent transport, we performed Single Particle orientation Tracking (SPoT) on long CCNTs, enabling simultaneous tracking of position in 2D and orientation in 3D[54–56] (Figure 5a). Notably, SPoT lacks the extended depth-of-focus capability of the double-helix SPT approach used above. As a result, in the 3D biophysical model where chambers and tunnels extend beyond the focal plane, recording long trajectories becomes challenging since particles may diffuse along the optical axis, thereby reducing the number of trajectories that can be reliably analyzed. Despite the limited number of trajectories accessible to SPoT, we obtained consistent and robust orientational signatures. More precisely, after segmentation of the trajectories, we retrieved 14 tunnels and 17 chambers. The orientational dynamics were quantitatively analyzed by calculating the Mean Square Angular Displacement (MSAD):

$$MSAD(\Delta t) = 4D_r\Delta t \tag{3}$$

The resulting values show a strong orientational hindrance in the tunnels ($D_r$= 0.4 rad$^2$/s), while in the chambers, CCNTs have a much higher rotational freedom in the chambers ($D_r$= 1.4 rad$^2$/s) (Figure 5b).

Furthermore, the persistence of an anisotropic particle's initial orientation quantified by the Reorientation Time correlation Function (RTF) was evaluated:

$$RTF(\Delta t) = Se^{\frac{-\Delta t}{\tau}} \tag{4}$$

where S is a constant, $\Delta t$ is the time lag, and $\tau$ gives the characteristic length of orientational memory.

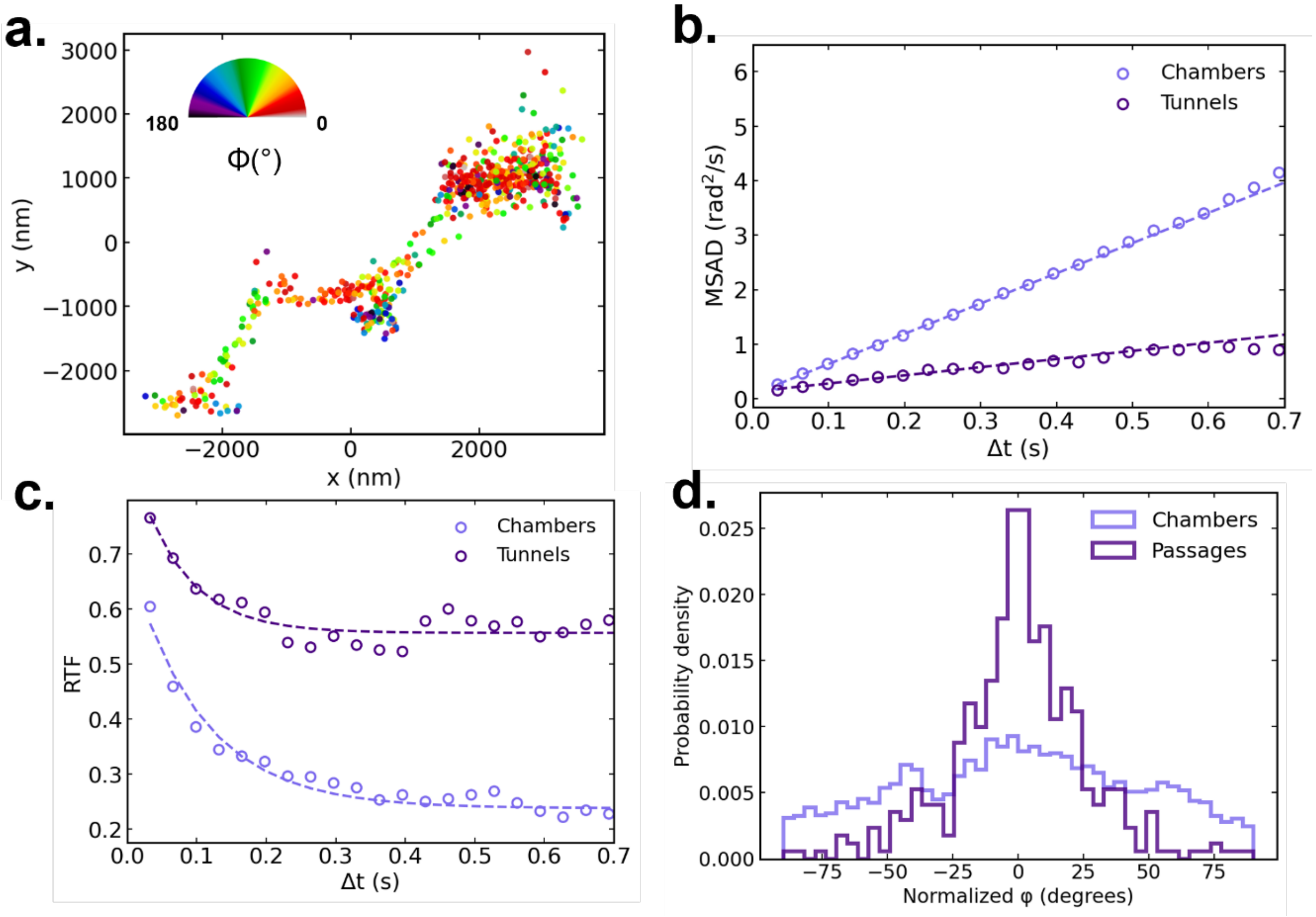

**Figure 5 Geometrically frustrated diffusion induces facilitated escape** a. The spatial distribution of the planar angle (Φ) of a representative trajectory of a CCNT exploring the chambers and tunnels in the biophysical model reveals spatial heterogeneities in particle orientation. b. The Mean Squared Angular Displacement (MSAD) further confirms the hindered rotation of the particles in the tunnels as opposed to the chamber exploration. c. The Reorientation Time correlation Function (RTF) shows longer memory of the initial orientation in the tunnels, while in the chambers, the particles quickly "forget" them. d. The normalized distribution of the angles in the chambers and tunnels reveals a strong orientational conformity in the tunnels. Whereas in the chambers, the particles explore the full spectrum of angles to orient themselves in.

The RTF shows that in the chambers, CCNTs almost completely lose "memory" of their initial orientation over time with a small residual plateau probably due to their length being comparable to the chamber dimensions, whereas in the tunnels, despite a slight initial loss of orientational memory, the correlation quickly plateaus without dropping further, indicating orientational persistence (Figure 5c). We next define a circularly normalized azimuthal orientation parameter to enable comparison across geometrically distinct regions. For each chamber or tunnel, the azimuthal angle distribution is recentered by subtracting its mean orientation, thereby setting the local average direction to zero. The angles are subsequently wrapped to a common interval (−90° to 90°), and the normalized distributions are aggregated across all regions as the Centered $\phi$. This pooled distribution highlights strong alignment in tunnels, in contrast to the nearly isotropic angular freedom observed in chambers (Figure 5d). Some slight preferential orientations are observed due to the mean CCNT lengths being of the order of the chamber dimensions. Similarly, the distribution of circularly normalized polar angles ("Centered $\theta$") displays nearly similar spread in both, the chambers and the tunnels (Supplementary Figure 4), confirming that only in-plane trajectories could be imaged and tracked. In contrast, for uCCNTs diffusing in the biophysical model, SPoT was unable to resolve the orientations of the particles. Given their small dimensions, uCCNTs can undergo fast rotational diffusion within the chamber, with a characteristic timescale of ~0.002 ms, much shorter than the camera integration time (33 ms) required for SPoT. As a result, their orientation averages out within a single exposure such that the recorded PSF corresponds to a time-averaged projection rather than a well resolved orientation PSF, precluding reliable extraction of particle orientation. Conversely, the characteristic timescale for free rotational diffusion of CCNTs is on the order of ~12 ms. Combined with the anomalous subdiffusion imposed by the biophysical model, an acquisition time of 33 ms is sufficient to capture and resolve particle orientations from the recorded PSFs. Analysis via SPoT indicates that CCNTs undergo rotational frustrations that correlate with transient trapping. Within this trapped state we observe a tendency for CCNT prealignment to the tunnels facilitating escape along these preferred orientations as indicated by a substantial lowering of the escape time scaling. In contrast, uCCNTs are characterized by rapid diffusion and frequent boundary interactions, suggesting escape dynamics based on efficient spatial sampling and stochastic exploration.

To further probe the geometric origin of this escape behavior, we tested whether it is governed by nanotube length relative to confinement size, rather than by either one alone, by repeating our measurements in a second, smaller droplet environment. Small droplets (SD, diameter 1.4μm (Supplementary Figure S5)) give rise to smaller chambers (SC, characteristic chamber length measured from trajectories, $D_c \approx 250$ nm) which we compare here with the larger chambers (LC) used throughout the rest of this work. As the two confinement conditions differ in volume by a factor of ~24 (Supplementary Figure 6a), we first rescaled the escape time by the characteristic chamber volume to allow direct comparison. However, normalized escape times remained distinct for nanotubes of identical length in the SC and LC conditions (Supplementary Figure 6b), indicating that confinement volume alone does not account for the observed escape dynamics.

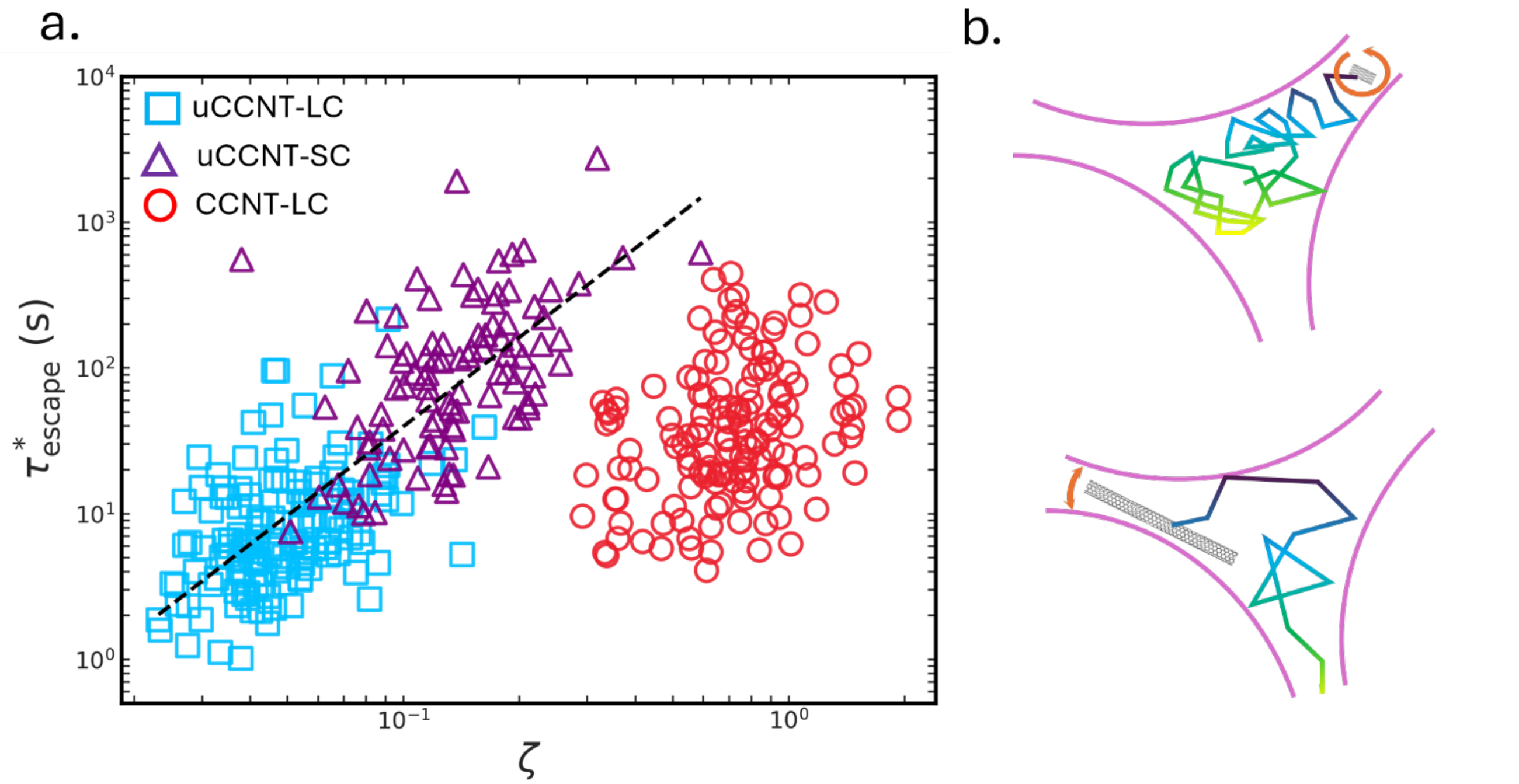

**Figure 6 Geometric scaling of nanotube escape dynamics (a)** Escape times rescaled by the characteristic confinement volume scale, $\tau^*_{esc} = \tau_{esc}/{D_C}^3$, are plotted as a function of the dimensionless nanotube-to-confinement size ratio, $\zeta = L/D_C$, for the different experimental conditions (LC and SC correspond to chambers in large and small droplets respectively). The data exhibit a common power-law scaling (exponent ≈ 2.13) over the low-$\zeta$ regime, while a systemic deviation emerges above a characteristic $\zeta$, marking a crossover to a distinct escape regime. **b.** Schematic representations of nanotube escaping from chambers through connecting tunnels. For uCCNTs, whose dimensions are small relative to the chambers and tunnels, the nanotubes retain substantial rotational freedom (top). In contrast, the CCNTs experience restricted rotational freedom and preferentially align with the available exit pathway, facilitating escape through the connecting tunnels (bottom).

We therefore considered instead the dimensionless ratio $\zeta = L/D_C$ which directly compares nanotube length to the chamber dimension. Plotting the volume-normalized escape time $\tau_{esc}^{*} = \tau_{\rm esc}/\rm V$ , where V represents the volume of the convex hull used to characterize the chamber trajectory, against $\zeta$ reveals a common power law scaling, $\tau_{\rm esc}{\sim}\zeta^{2.13}$ across the low $\zeta$ regime (Figure 6a). Despite spanning distinct ranges of $\zeta$, the uCCNT data from the SCs and LC conditions collapse onto this same scaling relationship, showing that the escape dynamics are governed by nanotube length relative to the chamber dimension rather than by the confinement volume or nanotube length alone. At larger $\zeta$, the CCNT-LC data systematically deviate above this scaling, marking a crossover to a distinct escape regime. This regime is consistent with the restricted rotational freedom and preferential tunnel-alignment of the longer CCNTs independently observed by SPoT microscopy (Figure 6b).

**Discussion**

Our work indicates that anisotropic particles with dimensions smaller than their surroundings exhibit rapid volumetric diffusion together with frequent boundary interactions, resulting in highly efficient spatial sampling. At short timescales, they exhibit rapid diffusion, while at longer times, interactions with spatial boundaries induce confinement-driven hindrance, resulting in anomalous diffusion.

Conversely, as particle lengths approach the characteristic dimensions of the environment, spatial confinement strongly impacts diffusion even at short timescales, while still allowing efficient motion. Across two confinement regimes differing in chamber volume by a factor of ~28, we find that the normalized escape time follows a common power-law scaling with the dimensionless length-to-confinement ratio $\zeta = L/D_C$ (exponent ≈ 2.13, low-ζ regime), rather than with nanotube length or chamber dimension independently. This directly confirms that it is this ratio, and not particle length alone, that governs escape dynamics. At larger ζ, however, the CCNT-LC data systematically depart from this scaling: escape from confined regions is instead facilitated by the probe geometry as frequent interactions with the boundaries hinder rotational diffusion and effectively pre-align the particles along available pathways, promoting funneling[57,58] and increasing the escape probability (i.e., reducing residence times) compared to a pore-like geometry. This effect is reflected in the weak dependence of escape times on nanotube length.

We further examine the theoretical basis for this weak length-dependence. We compare our escape-time measurements to a quantitative model derived by Doi and Xu[44]. This model describes the Brownian escape of 1D, needle-like objects through a narrow opening. It treats escape as a two-stage process: a free phase and a capture phase. In the free phase, the particle retains full translational and rotational freedom within the chamber; in the capture phase, one end of the particle enters the opening, imposing strong constraints on its rotational motion. For CCNTs, whose length is comparable to the chamber dimensions (Supplementary Figure 2b), escape is strongly dominated by the capture phase, so that their escape time ($\tau_{long}$) can be expressed as[44] :

$$\tau_{long} = \frac{2VL}{3\sqrt{3}a^2 D_t} \tag{5}$$

where $V$, $L$, $a$ , and $D_t$ represent the bounding volume, the nanotube length, the radius of the escape hole and the translational diffusion coefficient, respectively. For uCCNTs, whose length is much smaller than the chamber dimensions, escape is instead governed by the free-exploration regime:

$$\tau_{short} = \frac{V\left(1 + \frac{4aLD_r}{9D_t}\right)}{2\pi D_t a\left(1 + \frac{L^2 D_r}{6D_t}\right)} \tag{6}$$

with $D_r$, the rotational diffusion coefficient. Here $D_t$, and $D_r$ are given by[44]:

$$D_t = \frac{kT}{3\pi\eta L}\left(\log p + X_T(p)\right) \tag{7}$$

$$D_r = \frac{3kT}{\pi\eta L^3}\left(\ln p + X_R(p)\right) \tag{8}$$

where $k$, $T$, $\eta$ and $p$ represent the Boltzmann's constant, the absolute temperature, the viscosity of the fluid, and the aspect ratio of the anisotropic particle, respectively. The terms $X_T(p)$ and $X_R(p)$ are finite length correction terms.

While these expressions assume purely Brownian translational and rotational dynamics, they provide a useful geometric baseline for how escape times are expected to scale with particle length in the absence of tube-wall memory effects. To take those into account, we performed a Monte-Carlo parameter sweep simulation based on these equations and investigated the length dependent scaling of escape times. The values of particle lengths and escape-hole radii were drawn from our own measured distributions (Supplementary Figures 1, 2c, see Methods). We find that a tenfold increase in nanotube length predicts an approximately tenfold increase in escape time (median ratio 11.28 ± 9.66, Figure 4b inset). Hence, even purely Brownian theory predicts a linear length dependence on the scaling of escape times. This result strongly deviates from our experimental observation.

This deviation can be placed in the broader context of escape-time theories for confined diffusion. Geometry-controlled reaction kinetics frameworks predict that the mean first-passage time to a small target scales primarily with the ratio of confining volume to target size, with only weak (logarithmic) dependence on the diffusing particle's own dimensions once orientational degrees of freedom are accounted for[58,59]. Our results are qualitatively consistent with this picture, but go further by showing that anisotropy itself, through wall-induced pre-alignment, actively suppresses the residual length-dependence expected even from these frameworks.

This behavior also contrasts with escape-time scalings reported for spherical or near-isotropic nanoparticles escaping rigid porous cavities, where escape time typically scales more strongly with the size of the diffusing probe relative to the pore throat[29–31]. The comparatively weak length-dependence we observe for anisotropic 1D nanotubes therefore appears to be a distinct signature of orientational degrees of freedom coupling to escape kinetics, rather than a generic feature of confined diffusion in interconnected pore networks.

A relevant question is whether the softness of the emulsion interfaces, in contrast to solid porous media, affects the observed dynamics. To address this, we estimated the characteristic amplitude of thermally-driven interfacial fluctuations using classical capillary wave theory, $l_T$ = SQRT($k_B T/\gamma$). Using an interfacial tension representative of silicone-oil-in-water, SDS-stabilized emulsions such as used here[60] ($\gamma \approx 10$ mN/m), we obtain $l_T \approx 0.6$ nm at room temperature, and $l_T \approx 2$ nm even under a conservative order-of-magnitude lower bound for $\gamma$. This thermal roughness is one to two orders of magnitude smaller than the confinement length scales relevant to nanotube transport (tunnel radius 70-110 nm; chamber size 640-860 nm), and the conclusion is insensitive to the exact value of $\gamma$ within this range. We therefore conclude that, on the length scales governing nanotube confinement and escape, the droplet interfaces behave as effectively rigid boundaries, consistent with the faceted, polygonal droplet morphology observed under compression (Figure 1b, Supplementary Figure 2a) and with the absence of measurable coalescence or ripening over the imaging timescale.

This conclusion is independently corroborated by the analysis of nanotube dynamics. A mechanically soft, viscoelastic confinement is a classic physical origin of fractional Brownian motion (fBm) and related generalized-Langevin descriptions of anomalous diffusion[3,61], which would be expected to produce a scale-free, persistent VACF, a single spatially homogeneous anomalous exponent, and turning-angle statistics independent of tracer length. Chiefly, it would also show gaussian statistics, with a consistently low non-gaussianity parameter (NGP) converging to zero over long timescales[62]. None of these signatures is observed: the VACF (Figure 2b) relaxes to zero after a finite decorrelation time (1.03 s for CCNTs and 0.95 s for uCCNTs), the MSD of long trajectories (Figure 2d) resolves discrete dynamical regimes that map directly onto independently visualized chamber and tunnel geometries (Figure 2e, Figure 3a), turning-angle peak shifts systematically with nanotube length (110° for uCCNT vs. 140° for CCNT, Figure 2c) and the NGP of both uCCNTs and CCNTs are divergent with increasing timescales (Supplementary Figure 7). These are the classic signatures of obstructed diffusion by rigid-like boundaries, not of fBm[38]. These two independent lines of evidence show that our system therefore already operates in the geometrically "hard" regime relevant to nanotube transport, despite the physically deformable nature of the droplet. Systematically tuning the interfacial tension to explore a genuine soft-to-hard transition remains an interesting avenue for future work.

Our results reveal that geometric frustration, arising from repeated wall-mediated collisions within a continuous, structured environment, is key to understanding anomalous diffusion and length-dependent escape dynamics of a rigid 1D tracer. In the absence of pore-mediated trapping, wall deformation or binding interactions, the convex architecture of interconnected chambers and tunnels alone is sufficient to generate antipersistent and subdiffusive behavior of the nanotubes. This anomalous transport does not rely on strong trapping or persistent alignment, but instead emerges from interactions with a heterogeneous geometry, where boundary collisions and intermittent confinement progressively shape particle motion. Nanotube length dictates how this geometric frustration is resolved. Short nanotubes primarily undergo near-isotropic, Brownian-like exploration of the volume, with rotational diffusion that may only be transiently affected during collision events. Their escape is thus governed by a first-passage process, dependent on the probability of reaching an opening through stochastic search. In contrast, longer nanotubes, whose dimensions approach those of the confining chambers, experience sustained rotational constraints. Their extended geometry promotes alignment along available spatial pathways, effectively pre-aligning them within tunnels and facilitating directional transport. Overall, the interplay between particle anisotropy and local architecture can enhance, rather than hinder, transport, giving rise to geometry-driven funneling effects.

The weak length-dependence of escape times reported here is conceptually related to, yet mechanistically distinct from, the enhanced nanorod mobility documented in entangled and cross-linked polymer networks[40,41]. In those systems, rod reorientation facilitates hopping between adjacent, effectively concave mesh openings, and the resulting diffusion enhancement is often understood as a decoupling from local polymer relaxation. In our convex, wall-dominated architecture, no such mesh-hopping occurs; instead, repeated collisions with continuously curved boundaries progressively pre-align the nanotube with the escape pathway itself, so that the same rotational-translational coupling that speeds up mesh traversal in network systems here suppresses the length-dependence of escape from a single confining chamber. This distinction highlights that anisotropy-mediated enhancement of transport can arise from at least two distinct geometric mechanisms (mesh-hopping in concave network topologies, and wall-induced pre-alignment in convex, continuous landscapes) both converging on a weaker-than-expected dependence on particle length.

These confinement ratios can be placed in the context of related systems. In entangled polymer melts, nanorods diffuse anomalously fast once their length exceeds the entanglement mesh size while their diameter remains below it ($L/\xi \approx 5.4$, $d/\xi \approx 0.6$ for $TiO_2$ nanorods in polystyrene[40], a confinement ratio remarkably close in magnitude to $L/a_{tunnel}$ reported here for CCNTs, despite the absolute length scales differing by more than an order of magnitude). Within our own prior work on the brain extracellular space using the similar CNT probes, extracellular width was found to vary down to ~40 nm[37], which is directly comparable to the tunnel radii reported here (70-110 nm), supporting the relevance of this biophysical model as a geometric proxy for that biological system.

Beyond the specific system studied here, these findings may have broader implications for transport in heterogeneous environments. Biomolecular activity in continuous intra- or extracellular spaces (as opposed to pore mediated transport e.g. through nuclear pores), filtration through continuous porous materials, transport in crowded soft-matter systems, and heterogenous catalytic processes are all governed by hindered diffusion where confinement and topology can strongly modulate particle dynamics. By demonstrating that anomalous diffusion can be governed by funneling effects arising from the coupling between geometric constraints of the environment and particle anisotropy, we provide a physical framework for transport in complex systems, opening avenues for the design of materials and architectures with controlled diffusion properties.

## Materials and Methods

### *Droplet preparation*

The emulsion was synthetized by extrusion through Shirasu Porous Glass (SPG) membranes as follows: 5mL of silicone oil (viscosity = 50 cSt, Sigma Aldrich) was extruded into a solution of 10mM SDS through a cylindrical SPG membrane with a mean pore size of 1.6 micrometers, using an Internal Pressure Micro Kit from SPG Technonolgy Co., Ltd. The droplets were then stored at room temperature. Their average diameter was ~5 µm.

To prepare the biophysical model, 5 µm diameter emulsion droplets were labelled and index-matched by diluting them at a 1:4 volume ratio in a combined labelling and index-matching solution. The index-matching buffer consisted of a 10 mM solution of sodium dodecyl sulfate (SDS) in a 1:1 (v/v) mixture of water and glycerol. For fluorescent labelling, a saturated solution of Nile Red in the index-matching buffer was used. Depending on the desired labelling intensity, the Nile Red solution was further diluted in the index-matching buffer prior to use.

Emulsions were incubated in this solution for several days to allow complete phase separation between the emulsion and aqueous phases. Once phase separation was achieved, CCNTs were introduced by gently pipetting 1 µL of the CCNTs (diluted to 1:20,000 in the index matching buffer) directly into the emulsion phase only, taking care not to disrupt the phase boundary. To obtain a densely packed droplet system, the sample was centrifuged at 4500 g, facilitating close packing of the emulsions. This configuration allows the CCNTs to explore the extracellular regions between droplets. The final sample was carefully scooped onto a coverslip with an imaging spacer (50 µm thickness) for imaging with the spatula. A similar procedure was used to prepare packed droplets with uCCNTs.

The 2 µm diameter droplets were obtained through emulsification in a Couette apparatus as described by Brujić J, et al.[63]. Briefly, a coarse emulsion is first obtained by manually mixing 18 ml Silicone oil 50 cSt in 30 ml aqueous phase containing 3% w/w Tergitol (Type 15-S-7, Sigma-Aldrich) and 3% w/w alginate (Alginic acid from brown algae, Sigma-Aldrich). This solution was then injected inside a narrow gap Couette mixer (gap size = 100 µm) and sheared at 50 rpm. The resulting emulsion was finally washed three times in 500 ml of aqueous solution containing 10 mM SDS, using a separating funnel.

***Droplet Imaging***

Prepared biophysical model samples containing CCNTs were imaged using a Nikon Eclipse Ti inverted microscope equipped with a Nikon Plan Apo 60× IR water-immersion objective (NA 1.27) and a double-helix phase mask (Fig 1a). White-light imaging was used to visualize the overall droplet morphology (Fig 1b).

For fluorescence imaging of the biophysical model environment, excitation was performed using a 568 nm laser at an intensity of 0.14 kW/cm$^2$ at the focus of the objective. Emission was collected using a 655 nm long-pass filter and recorded with a Teledyne ProEM electron-multiplying CCD camera. Optical sectioning was achieved using the Sparq module (Bliq Photonics), which enabled axial resolution down to ~500 nm.

For the SD, the characteristic chamber dimensions (~250 nm median) approached the diffraction-limited length scale, precluding reliable resolution of the chamber-tunnel geometry by fluorescence/HiLo imaging. Droplet sizes were therefore determined from white light images. From the uCCNT trajectories, chambers could be identified based on the trajectory geometry; however, uCCNTs typically spent only 3-5 time points within the tunnels, making their segmentation and quantitative characterization unreliable. Consequently, only the chamber regions were segmented and analyzed for the smaller droplets.

***3D SPT imaging and localization***

Videos of diffusing (6,5) CCNTs and uCCNTs were acquired in the near-infrared range by exciting at 985 nm (AeroDIODE) with an intensity of ~4 kW/cm$^2$ and detecting emission through a 1100 nm long-pass filter (Fig 1c). A double helix phase mask was used for engineering the PSF to localize in the lateral and axial directions. To maintain uniform excitation regardless of nanotube orientation, the laser beam was circularly polarized. Near-infrared acquisitions were made possible with the use of a C-RED2 InGaAs camera (First

Light Imaging) at frame rate of 30Hz. Localization of single particles in 3D based on the double helix PSF was achieved using the ZOLA plugin on Fiji. First, a calibration of the phase mask was performed by acquiring a Z-stack of stationary nanotubes around the focal plane, with 50 nm steps. At each step, 50 images were recorded and averaged before moving to the next step. The full range of PSFs was monitored by acquiring images in the range -3 µm to +3 µm. The ZOLA plugin on Fiji was used for phase retrieval of the double helix phase mask and to calibrate it for further localizing single particle trajectories[49]. Linking of these localizations into trajectories were carried out using home-built python codes using the trackpy library[64].

***SPoT imaging and localization***

CCNTs orientation were detected using SPoT, a single-particle tracking epi-fluorescence microscopy technique based on dipole spread function (DSF) engineering[33] . A vortex half-wave plate placed at the objective's back focal plane converts radially and azimuthally polarized components of the emitted fluorescence into two orthogonal linear polarization states. These components are then separated by a polarizing beam splitter, producing two spatially distinct images that are projected side-by-side onto an InGaAs camera. The 3D orientations and in-plane localization of single CCNTs were retrieved using the deep learning technique Deep-SMOLM[65] as described in details in Ruan et. al[33]. In short, the neural network was trained on a dataset of 2000 simulated images. Each training image contained 5-10 CCNTs with randomized 2D position, 3D orientations and emitter intensities with signal to noise ratios matching experimental conditions. The focal plane position was varied over a range of z ~ [-500nm, 500nm].

***Trajectory Analysis***

Diffusion analysis of MSD, MSAD, VACF, distribution of turning angles, RTF and normalized orientations of centered azimuthal and polar angles were carried out with home-built python codes. The trajectories were segmented manually into chambers and tunnels using the CloudCompare software based on the local geometry of the point cloud: chambers were identified as regions where the trajectory point cloud extended comparably in all directions (quasi-isotropic, compact convex hull), whereas tunnels were identified as elongated, quasi-one-dimensional segments with a large aspect ratio between their principal and secondary extents. Segment boundaries were placed at the visually identified transition between these two point-cloud geometries, consistent with the independent chamber and tunnel dimensions obtained from HiLo images of the droplet packing (Supplementary Figure 2).

***Escape-time extraction***

Escape (residency) times reported in Figure 4a were extracted from approximately 80-90 independent long trajectories per nanotube population ($N_{CCNT}$ = 82, / $N_{uCCNT}$= 91). Within each trajectory, every complete chamber residency (defined as a continuous dwell beginning with the particle's entry into a chamber from a connecting tunnel and ending with its exit into a - possibly different - connecting tunnel) was retained as one independent residency-time sample, with escape time taken as the elapsed time between these two events. A single trajectory could therefore contribute more than one residency event. Truncated dwells, i.e. those for which recording began or ended while the particle was already inside a chamber, were excluded from the analysis. No explicit chamber-size selection criterion was applied meaning that all valid, non-truncated chambers were retained, yielding the final pool of 150 (CCNT) and 148 (uCCNT) residency-time events reported in the manuscript. Similarly for the uCCNTs diffusing in the SD, 97 non truncated chambers were extracted. The resulting distributions of chamber sizes for the two nanotube populations, which overlap substantially, are shown in Supplementary Figure 2c.

***Quantification of local geometry***

To independently quantify the dimensions of the chambers and tunnels, optically sectioned images of the biophysical model obtained by HiLo were analyzed using Fiji. A threshold was applied to segment the chambers, and their dimensions were estimated by the diameter of the inscribed circle into each chamber (Supplementary Figure 2), resulting in a mean chamber dimension of 760 ± 170 nm. These values are consistent with those obtained from segmented trajectories, where a convex hull was fitted to each trajectory to estimate an effective diameter, defined as twice the distance from the centroid to the nearest point on the hull. For the long CCNTs, this resulted in a mean chamber dimension of 640 ± 290 nm whereas for the uCCNTs, the mean chamber dimension was found to be 860 ± 330 nm (Supplementary Figure 2b). For the tunnels, due to the limited resolution of HiLo images, their radii were determined only from trajectories. The radius was calculated as the geometric mean of the two smaller semi-minor axes from an ellipsoid fit onto each tunnel trajectory[36]. The mean radius of the tunnels from long CCNTs was determined to be 70 ± 20 nm while the uCCNT trajectories gave 110 ± 30 nm (Supplementary Figure 2c). The differences in estimated dimensions obtained from HiLo imaging and from nanotube trajectories arise intrinsically from the nature of the measurement techniques. SPT relies on localizing particles via their center of mass and therefore does not account for their full spatial extent or the steric constraints they experience. Due to their larger length, CCNTs interact more frequently with confining boundaries and experience stronger geometric hindrance than uCCNTs. As a result, trajectories of longer CCNTs tend to underestimate the true dimensions of the environment.

### *Simulations*

Escape-time ratios between long and short nanotubes were estimated using a Monte Carlo parameter sampling of the analytical model derived from equations 3, 4, 5, and 6:

$$\frac{\tau_{\mathrm{long}}}{\tau_{\mathrm{short}}} \approx \frac{\left(L_{\mathrm{long}} \ln(2L_{\mathrm{short}}/d)\right)(L_{\mathrm{long}}/a)}{L_{\mathrm{short}} \ln(2L_{\mathrm{long}}/d)(1+4L_{\mathrm{short}}/a)} \tag{9}$$

where $L_{\mathrm{long}}$ and $L_{\mathrm{short}}$ are nanotube lengths for the CCNTs and uCCNTs respectively, $a$ is the escape hole radius determined as the radius of the tunnels as observed by the uCCNTs, and $d$ is the nanotube diameter. Parameters were randomly sampled from experimentally constrained ranges ($L_{\mathrm{long}} = 200\text{–}1000$ nm, $L_{\mathrm{short}} = 30\text{–}150$nm, $a = 70\text{–}270$nm), with distributions centered on their respective mean values. The diameter was fixed at $d = 3$nm. For each realization, the escape-time ratio was computed, and a total of $N = 10000$ independent samples were generated.

**Acknowledgments**

The authors thank Jacques Fattaccioli for the use of the Couette apparatus. This work received financial support from the European Research Council Synergy grant (951294), Agence Nationale de la Recherche (ANR-24-CE09-2351-01), EUR Light&T (PIA3 Program, ANR-17-EURE-0027), the France-Bioimaging National Infrastructure (ANR-10-INBS- 04-01), and the Idex Bordeaux (Grand Research Program GPR LIGHT). LR is financially supported by the China Scholarship Council (Grant No. 202208330010). L.L.P acknowledges financial support from Agence Nationale de la Recherche (ANR-23-CE13-0025-02, ANR-23-CE13-0023-02, ANR-25-CE30-2977-01).

**Supporting information**

Quantification of droplet structure; Nanotube length to chamber and tunnel size ratios; Distribution of lengths of CCNTs and uCCNTs; Estimation of dimensions of the biophysical model; Diffusion analysis in bulk liquid ; Distribution of normalized polar angles; Distribution of normalized escape; Non-gaussianity confirms obstructed diffusion.

# Supplementary Information

# Orientational frustration drives enhanced diffusion of anisotropic particles in a liquid labyrinth

Rohit Mangalwedhekar[1,2], Limeng Ruan[1,2], Somen Nandi[1,2], Quentin Gresil[1,2], Marc Tondusson[1,2], Stephane Bancelin[1,2], Lea-Laetitia Pontani[3,4], Laurent Cognet[1,2,*]

*[1]Laboratoire Photonique Numerique et Nanosciences, Université de Bordeaux, 33400 Talence, France,*
*[2]LP2N, Institut d'Optique Graduate School, CNRS UMR 5298, 33400 Talence, France*
*[3] Sorbonne Université, CNRS, Laboratoire Jean Perrin, LJP, F-75005 Paris, France*
*[4] Sorbonne Université, CNRS, Inserm, Institut de Biologie Paris-Seine, IBPS, F75005 Paris, France*

**Table 1a: Quantification of droplet structure.**

| **Quantity** | **Population / method** | **Value** |
|---|---|---|
| Droplet diameter | HiLo imaging, (particle-independent) | 4.3 ± 1 µm |
| Chamber dimension $D_c$ | HiLo imaging (particle-independent) | 760 ± 170 nm |
| | CCNT-trajectory convex hull | 640 ± 290 nm |
| | uCCNT-trajectory convex hull | 860 ± 330 nm |
| Tunnel (constriction) radius a | CCNT-trajectory ellipsoid fit | 70 ± 20 nm |
| | uCCNT-trajectory ellipsoid fit | 110 ± 30 nm |
| Nanotube length L | CCNT (AFM) | 500 ± 200 nm |
| | uCCNT (AFM) | 50 ± 60 nm |
| Interstitial area fraction (2D) / Volume porosity ϕ (3D) | From HiLo imaging, particle-independent | 12 ± 1 % |

**Table 1b: Nanotube length to chamber and tunnel size ratios.** Both ratios are computed using the full range of chamber (640-860 nm) and tunnel-radius (70-110 nm) estimates in Table 1a (i.e. spanning the CCNT-derived, uCCNT-derived, and (for the chamber) independent HiLo values). The independent, particle-agnostic HiLo chamber estimate (760 nm) falls within the stated range in both cases.

| **Dimensionless ratio** | **CCNT** | **uCCNT** |
|---|---|---|
| $L / D_c$ | 0.58 - 0.78 | 0.06 - 0.08 |
| L / a | 4.5 - 7.1 | 0.45 - 0.71 |

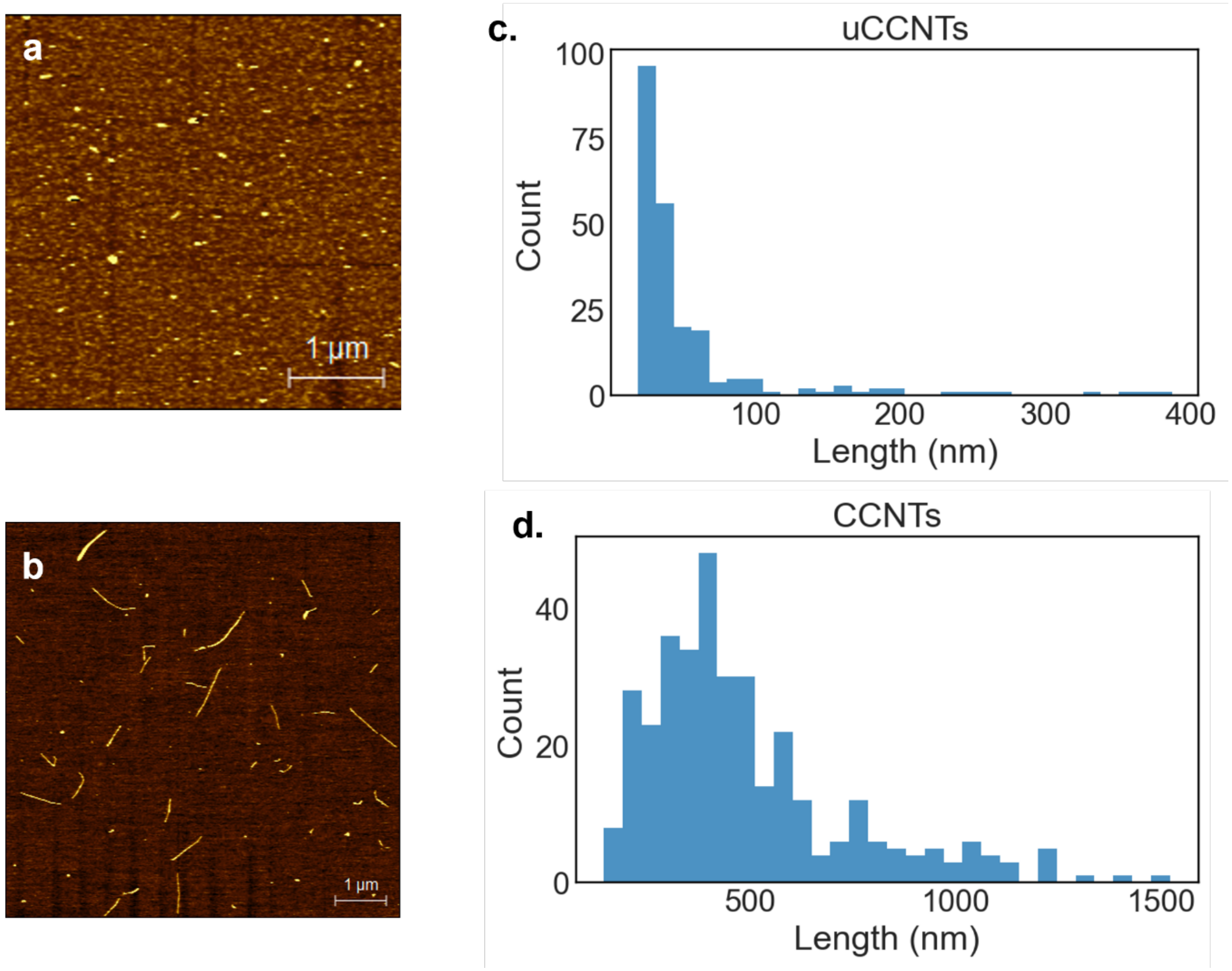

**Figure S1 Distribution of lengths of CCNTs and uCCNTs** Representative AFM images of a. uCCNTs and b. of CCNTs. Corresponding distribution of lengths of c. uCCNTs (mean length: 50 ± 60 nm) and d. CCNTs (mean length : 500 ± 200 nm), respectively.

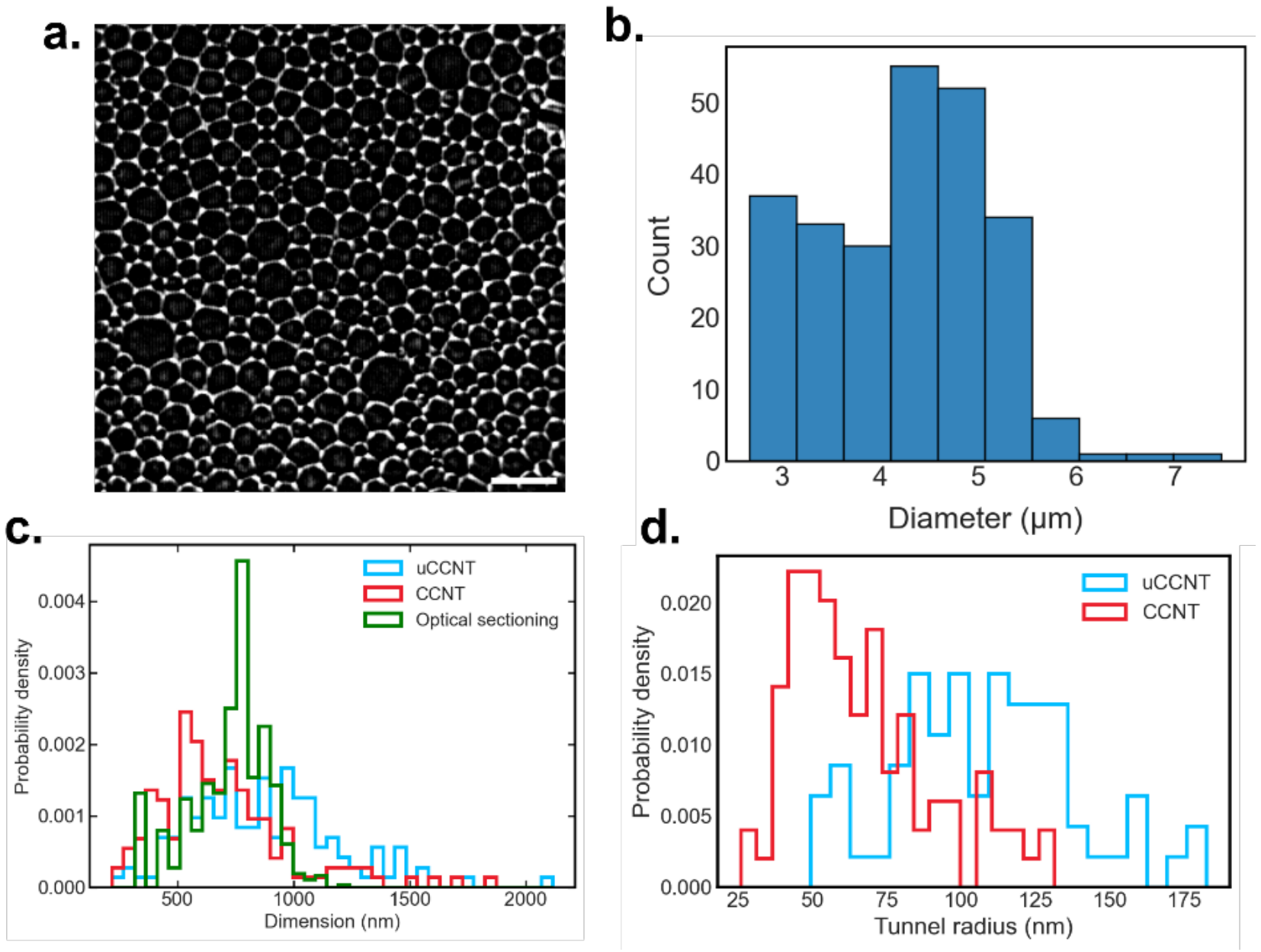

**Figure S2 Estimation of dimensions of the biophysical model** a. Representative optically sectioned fluorescence (HiLo) image of packed emulsion droplets (scale bar : 5 µm) b.The distribution of diameters of the droplets as measured by Hough circle transform on HiLo images using Fiji. The median diameter was 4.3 ± 1 µm. c. The distribution of dimensions of the chambers shows a median dimension of 640 ± 290 nm when measured from CCNT trajectories, 860± 330 nm from uCCNT trajectories, and 760 ± 170 nm from segmentation of the HiLo images. d. The distribution of tunnel radii estimated from the geometric mean of fitted ellipsoids provides a median radius of 70 ± 20 nm with the CCNT trajectories and 110 ± 30 nm with the uCCNT trajectories.

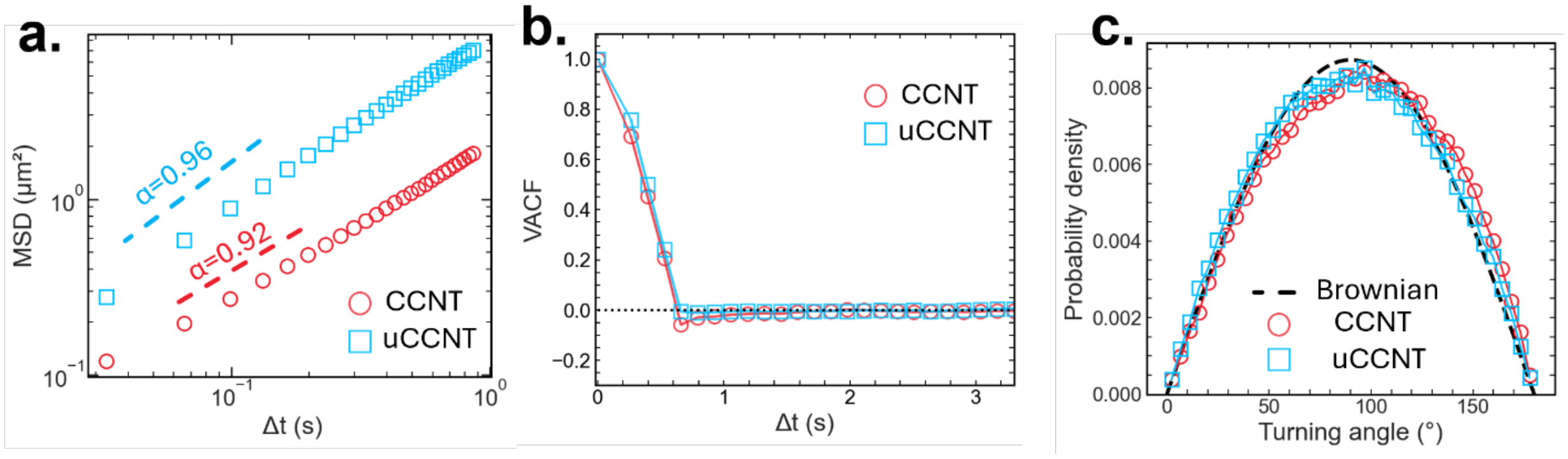

**Figure S3 Diffusion analysis in bulk liquid** a. MSD of CCNTs and uCCNTs in bulk liquid shows near Brownian diffusion for both CCNTs and uCCNTs. b. VACF of CCNTs and uCCNTs shows completely uncorrelated steps in Brownian motion in bulk liquid. c. The 3D distribution of angles closely follows the theoretical Brownian motion prediction (dashed black line).

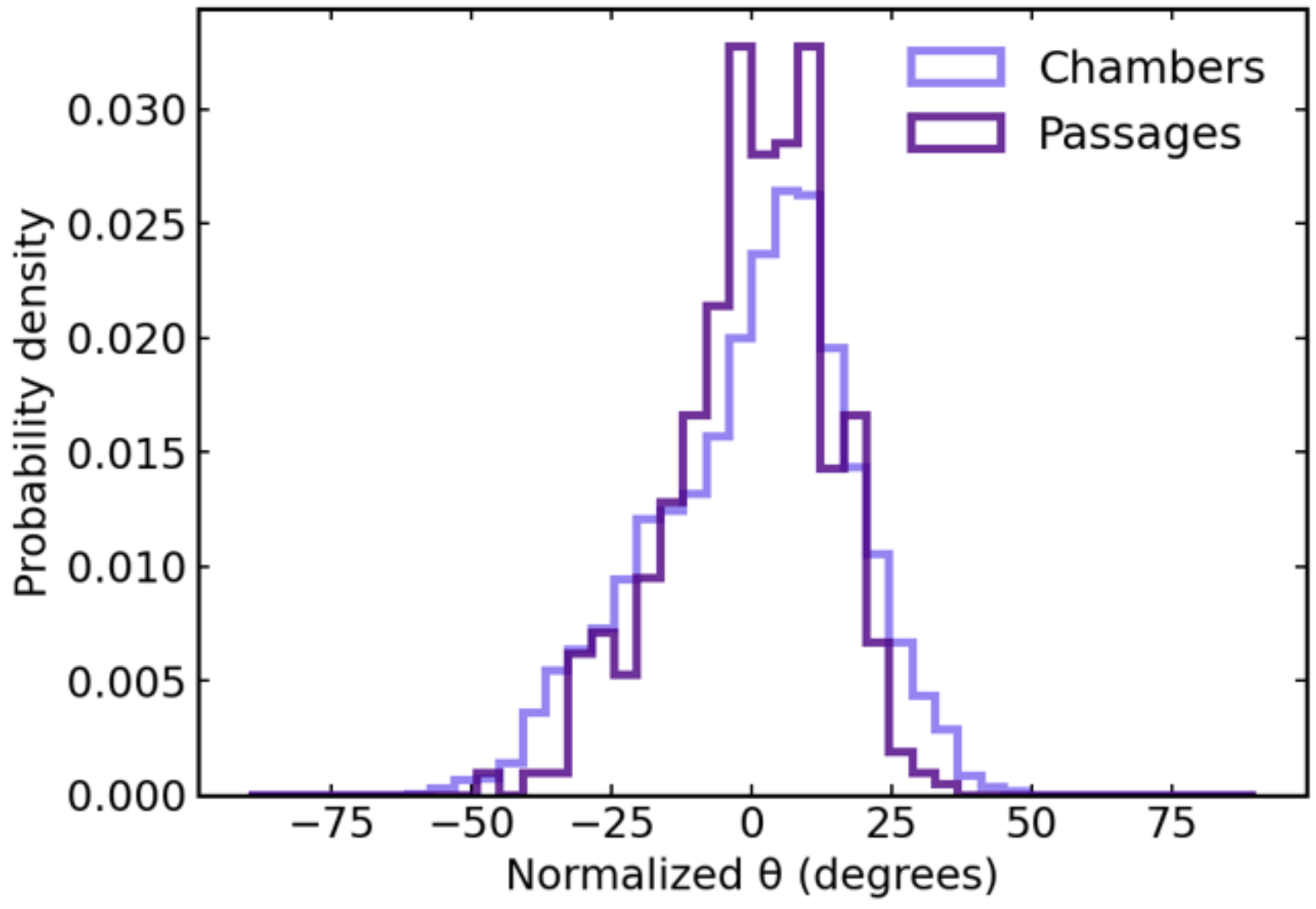

**Figure S4 Distribution of normalized polar angles** The normalized distribution of the polar angles in the chambers and tunnels reveals similar variance in both geometries since only in-plane trajectories can be imaged, tracked, and segmented.

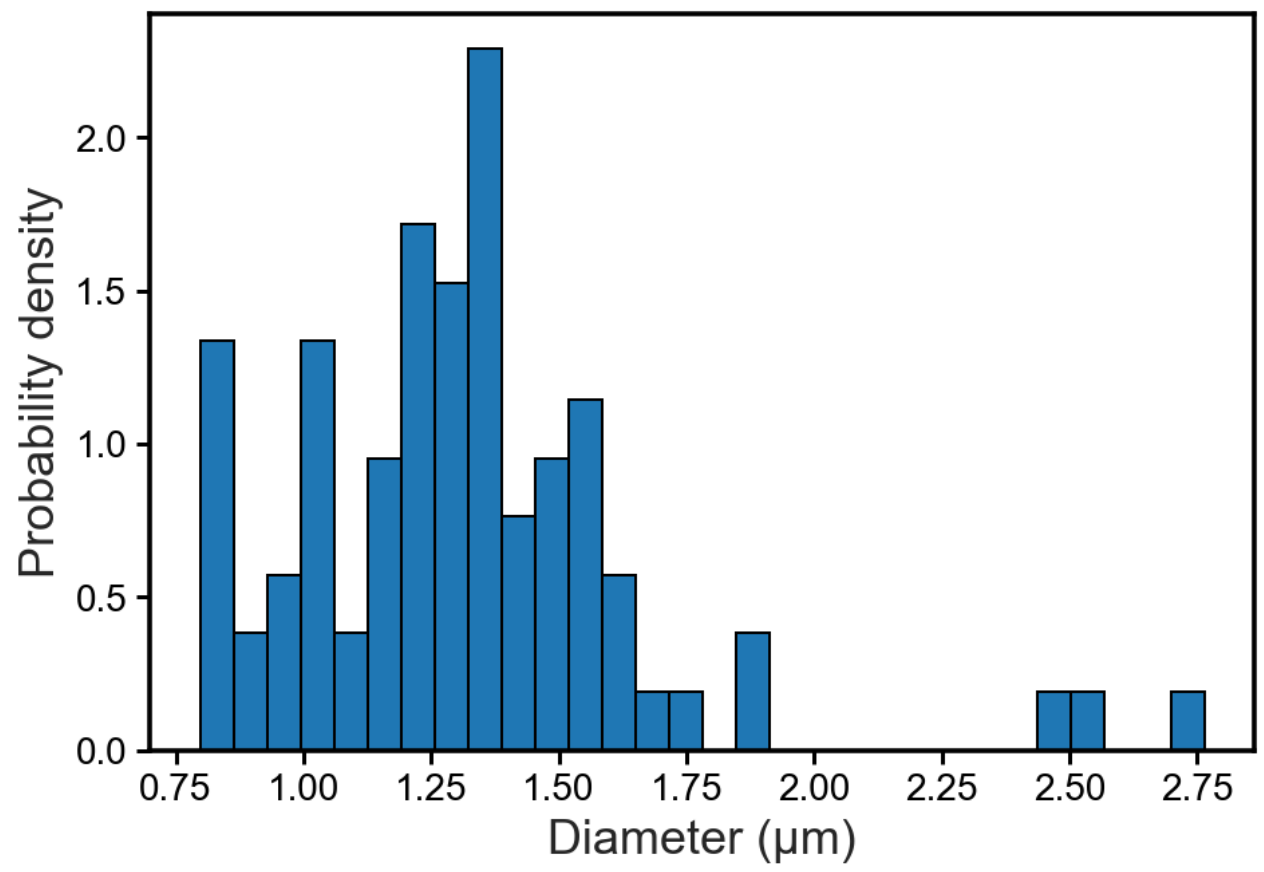

**Figure S5 Size distribution of small droplets (SDs).** Diameter distribution of SDs determined by threshold-based analysis of white-light microscopy images, yielding a mean diameter of 1.4 ± 0.5 μm.

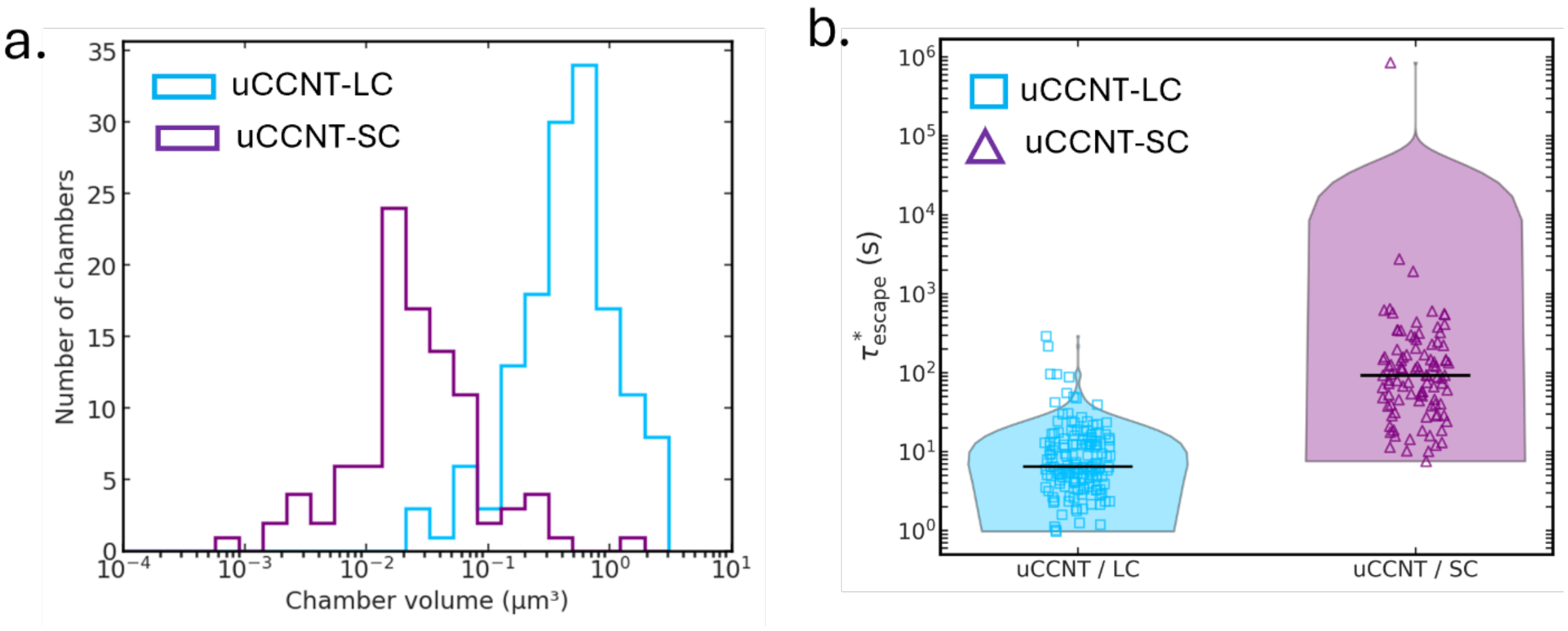

**Figure S6 Distribution of normalized escape times** a. Distribution of volumes of SC and LC obtained from diffusing uCCNTs b. Escape times of uCCNTs normalized by the characteristic volume scale. Despite this normalization, the escape times remain markedly different between LCs (median 7.28 s) and SCs (median 100.3 s).

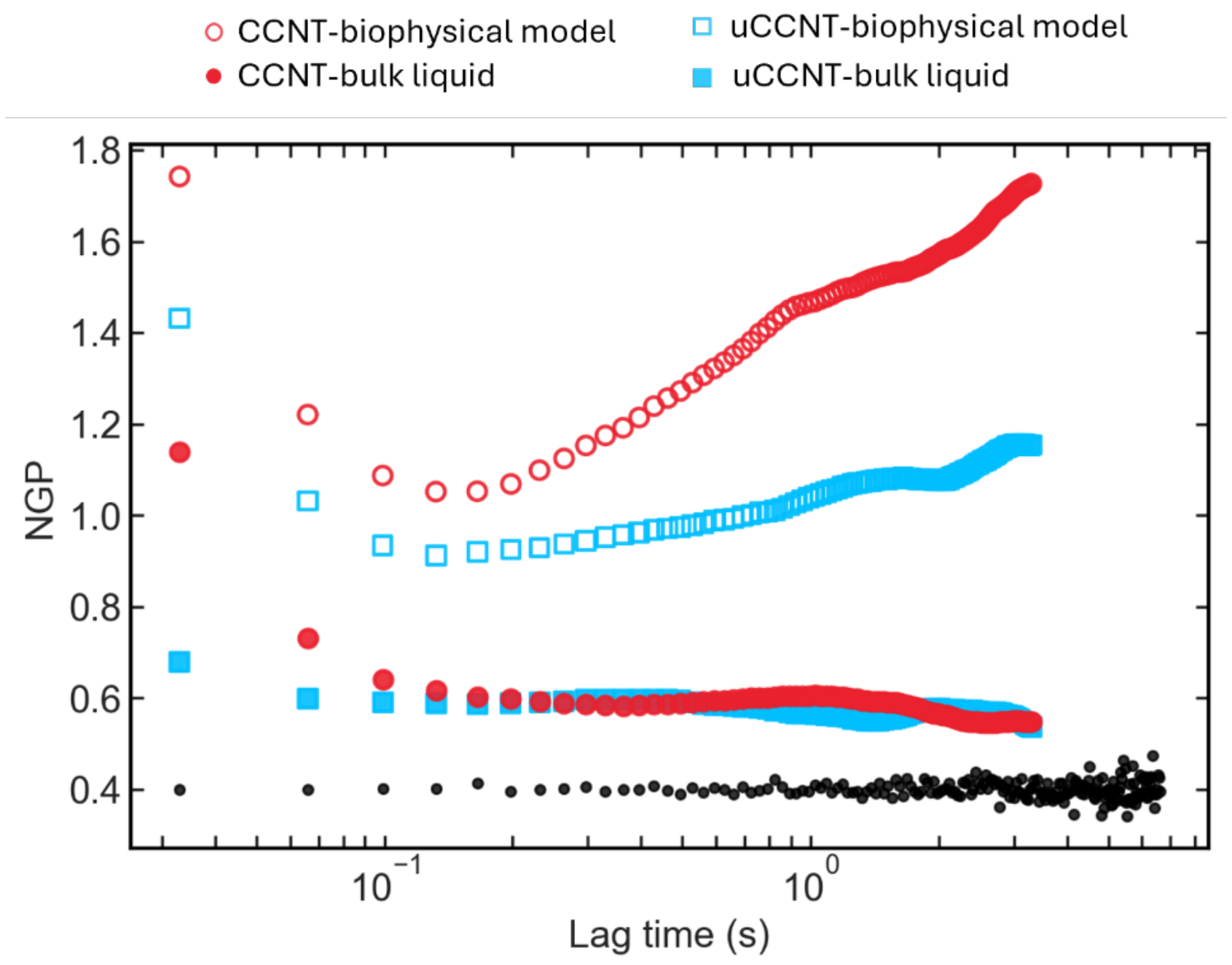

**Figure S7 Non-gaussianity confirms obstructed diffusion** The uCCNTs and CCNTs diffusing in the biophysical model show strong divergence with increased non gaussianity. The nanotubes diffusing in bulk Brownian fluid converge towards gaussianity. The NGP of a simulated fraction Brownian system has been shown as a reference (black).